%% file: main.tex
\documentclass[12pt, a4paper, logo, copyright, nonumbering]{xds}

\input{Meta/package}
\input{Meta/macro}

\usepackage{import}
\usepackage{mathtools}
\usepackage[ruled,vlined]{algorithm2e}

\SetCommentSty{mycommfont}
\SetKwProg{Fn}{Function}{:}{}
\SetKw{Continue}{continue}
\SetKw{KwAttribute}{Attribute}

\usepackage{hyperref}

\newcommand{\paper}{HFX}
\renewcommand{\textsc}[1]{#1}
\renewcommand{\emph}[1]{\textit{#1}} % Force emph to use textit

\title{\vspace{-0.2in}\centering \paper{}: Joint Design of Algorithms and Systems for Multi-SLO Serving and Fast Scaling}

\author[1,3,*]{Zahra Yousefijamarani}
\author[1,3,*]{Xinglu Wang}
\author[1,*]{Qian Wang}
\author[1]{Morgan Lindsay Heisler}
\author[1]{Taha Shabani}
\author[1,4]{Niloofar Gholipour}
\author[1]{Parham Yassini}
\author[1]{Hong Chang}
\author[2]{Kan Chen}
\author[2]{Qiantao Zhang}
\author[2]{Xiaolong Bai}
\author[3]{Jiannan Wang}
\author[1]{Ying Xiong}
\author[1,$\dagger$]{Yong Zhang}
\author[1,$\dagger$]{Zhenan Fan}

\affil[1]{Huawei Technologies Canada Co., Ltd.}
\affil[2]{Huawei Technologies Co., Ltd.}
\affil[3]{Simon Fraser University}
\affil[4]{École de technologie supérieure}

\renewcommand\Affilfont{\small}

\makeatletter
\renewcommand{\maketitle}{%
    \bgroup
    \setlength{\parindent}{0pt}
    \begin{adjustwidth}{0pt}{0pt}  
        \begin{flushleft}
            {\raggedright \titlefont \@title\par}%
            \vskip2pt
            {\raggedright \@author\par}%
        \end{flushleft}
    \end{adjustwidth}
    \vskip2pt

    {\abscontent}

    \begingroup
    \renewcommand{\thefootnote}{\fnsymbol{footnote}} % 使用符号脚注
    \footnotetext[1]{Equal contribution.}
    \footnotetext[2]{Corresponding authors: Yong Zhang <yong.zhang3@huawei.com>, Zhenan Fan <zhenan.fan1@huawei.com>.}
    \endgroup

    \thispagestyle{firststyle} % 保留你原来的第一页页眉/页脚
    \egroup
}
\makeatother

\begin{abstract}

Large language model (LLM) serving faces the dual challenge of meeting strict user-specific service-level objectives (SLOs) while minimizing computational cost under dynamic, multi-task workloads. Existing approaches either rely on static scheduling policies or focus on single-task settings, limiting their applicability in real-world deployments with heterogeneous requests, variable prompt lengths, and elastic scaling requirements.
We present \paper{}, a production LLM serving system that jointly optimizes request scheduling and elastic scaling across model replicas to satisfy diverse SLOs. \paper{} introduces a \textbf{scheduler} that performs proactive budget estimation and prioritization to ensure SLO compliance for both new and in-flight requests. \paper{} also integrates a \textbf{scaler} that supports fast device-to-device (D2D) weight transfer, reducing cold-start latency. Additionally, the system supports both colocated and disaggregated prefill/decode deployments, enabling adaptation to diverse workload patterns and cloud environments.

Through extensive experiments on multi-task workloads, we demonstrate consistently higher SLO attainment, lower end-to-end latency, and lower NPU usage cost by up to 4.44×, 65.82\%, and 49.81\%, respectively, compared to state-of-the-art systems. Our results highlight the effectiveness of SLO-aware scheduling and scaling in practical LLM serving, providing a robust framework for cost-efficient and SLO-compliant deployments.

\end{abstract}


\begin{document}
% \begin{CJK*}{UTF8}{gbsn}

\maketitle

\newpage
\tableofcontents
\newpage

%-------------------------------------------------------------------------------
\section{Introduction}

As large language models (LLMs) \cite{grattafiori2024llama3herdmodels,achiam2023gpt,deepseekai2024deepseekv3technicalreport, team2024gemma, touvron2023llama,yang2024qwen2.5} become integral to enterprise and consumer applications deployed on Huawei Cloud, maintaining predictable performance under diverse service-level objectives (SLOs) has become a central operational challenge. Production environments exhibit substantial variation in request size, priority, and latency sensitivity, making static scheduling and resource allocation insufficient. Without careful coordination, latency-critical requests may be delayed by long or low-priority tasks, resulting in SLO violations and increased cost.

% As large language models (LLMs) \cite{grattafiori2024llama3herdmodels,achiam2023gpt,deepseekai2024deepseekv3technicalreport, team2024gemma, touvron2023llama,yang2024qwen2.5} are increasingly deployed through Huawei Cloud to power enterprise and consumer-facing applications, including real-time assistants, code generation tools, and large-scale document summarization \cite{achiam2023gpt,wu2023less,biester2024llmclean,jiang2024survey, van2023clinical}, maintaining predictable performance under diverse service-level objectives (SLOs) has become a critical operational challenge. Production workloads exhibit wide variability in request size, priority, and latency sensitivity, making static scheduling and resource allocation economically insufficient . Without careful management, latency-critical requests can be delayed while low-priority tasks overconsume resources, resulting in SLO violations and increased operational cost.

% \textcolor{red}{Niloofar:"economically inefficient" can be replaced by insufficient} Applied

A central bottleneck in production clusters is the resource arbitration for multi-SLO workloads, where requests impose distinct requirements for Time-to-First-Token (TTFT), Time-Per-Output-Token (TPOT), or throughput. For example, interactive chatbots require low TTFT for responsive user interactions, whereas batch summarization jobs can tolerate longer prefill times but demand high decode throughput. In production clusters, these workloads must be co-scheduled across shared hardware, with careful balancing to prevent high-priority tasks from being blocked or low-priority tasks from monopolizing resources. This heterogeneity makes scheduling decisions fundamentally more complex than in single-SLO or single-task settings.

% \textcolor{red}{Niloofar slightly different to keep the industrial tone:A central bottleneck in production clusters is the resource arbitration for multi-SLO workloads, where requests impose conflicting demands for Time-to-First-Token (TTFT), Time-Per-Output-Token (TPOT), and aggregate throughput. For example, interactive applications demand aggressive TTFT for responsiveness, whereas batch summarization jobs tolerate startup delays but require sustained decode throughput. In shared multi-tenant environments, these workloads must be co-scheduled with precise isolation to prevent high-priority traffic from being blocked by compute-bound prefill bursts or low-priority tasks from monopolizing memory bandwidth.} Thanks Applied partially( not just by compute-bound prefill )

Even with effective scheduling, sustaining SLO compliance further depends on timely elastic scaling. However, conventional auto-scaling mechanisms struggle to provide timely capacity adjustments because loading large model weights introduces substantial cold-start delays. To avoid these delays, operators often over-provision resources, trading cost efficiency for minimal SLO risk.

% \textcolor{red}{Niloofar industrial tone: Elastic scalability is crucial for maintaining SLO compliance during traffic volatility. While rapid, fine-grained scaling allows resources to adapt to sudden demand spikes, achieving this in production is hindered by the I/O latency of loading massive model weights from storage. Standard auto-scaling mechanisms often incur unacceptable cold-start delays, forcing operators to over-provision resources to buffer against load fluctuations.} Applied

Another key design consideration is the deployment mode of the Prefill/Decode pipeline. Colocated deployments optimize throughput by sharing hardware resources across prefill and decode, but make it difficult to meet strict TTFT and TPOT requirements, whereas disaggregated deployments allocate dedicated resources to each stage to satisfy tight latency SLOs, often at the cost of using more capacity \cite{hu2024inferenceinterferencedisaggregatellm, zhong2024distserve, wang2025prefill}. In practice, both modes are used in production, and the choice between them is often determined by workload composition and operational constraints rather than fixed task categories. Since production workloads mix latency-sensitive, throughput-oriented, and long-context tasks, supporting both modes allows operators to select the architecture best aligned with their operational constraints.

% \textcolor{red}{Niloofar, slight modification for industrial tone: Another key infrastructure decision is the deployment mode of the Prefill/Decode pipeline. Colocated deployments minimize TTFT by avoiding KV-cache transfer overheads, ideal for latency-sensitive tasks. Disaggregated deployments, conversely, isolate memory-bound decoding from compute-bound prefill interference, significantly improving goodput for long-context requests \cite{hu2024inferenceinterferencedisaggregatellm, zhong2024distserve, wang2025prefill}. Since real-world traffic mixes requests of diverse lengths and priorities, a rigid commitment to a single architecture limits operational efficiency. Supporting both modes allows operators to dynamically align the architecture with real-time workload characteristics.}    Applied partially. (we do not support dynamic changing so last sentence can not be applied,  minimize TTFT is also wrong

Existing scheduling and scaling approaches  \cite{nie2024aladdin, wu2025arrowadaptiveschedulingmechanisms, zhu2025polyserveefficientmultisloserving, cui2025optimizingsloorientedllmserving, zhang2025tempoapplicationawarellmserving, tang2025scorpioservingrightrequests, gu2025hasgpuefficienthybridautoscaling, 298766} either target at single-task workloads or limited to single-instance deployment.
These systems provide valuable insights but do not fully address the complexities of heterogeneous, multi-stage, and dynamically varying cloud workloads, leaving SLO attainment and cost efficiency under-optimized.
% \textcolor{red}{Niloofar: "theoretically valuable" maybe better instead of "they provide useful insights"}  not fair

To address these operational challenges, we introduce \textbf{\paper{}}, a novel unified LLM serving system for production-scale cloud deployments. \paper{} integrates algorithmic and system-level innovations to jointly optimize scheduling and scaling across colocated and disaggregated P/D execution, as shown in Figure \ref{fig:overview}. It is designed for real-world operational requirements, ensuring high SLO compliance, low latency, and cost-effective resource utilization.

% Applied Niloofar's version

% Figure \ref{fig:overview} presents an overview of the \paper{} system architecture. Incoming requests are first processed by the Orchestrator, which serves as the central control plane. It integrates the dispatcher, migrator, monitor, and scaler to jointly execute the system's core innovation: proactive, SLO-aware scheduling and scaling decisions. It routes the prefill and decode stages to the appropriate instance instances during P/D disagregated mode. Furthermore, the system is designed for closed-loop adaptation: system statistics and SLO feedback are continuously aggregated, enabling robust operation under dynamic load conditions and supporting both deployment modes.

Our system makes the following contributions:

\begin{itemize}
\item \textbf{Multi-SLO-Aware Scheduling:} We design a dispatcher that organizes requests into SLO-aware queues and assigns them to instances via an Instance-Priority Queue. In disaggregated P/D deployments, prefill and decode stages are scheduled independently using the Dispatcher and Migrator, ensuring each stage meets its SLO without interference, even under heterogeneous workloads.

\item \textbf{Elastic Scaling Framework:} A unified scaling controller dynamically adjusts resources based on load and utilization across both execution deployments, and supports prefill–decode role transitions in P/D mode. Device-to-device (D2D) weight transfer further cuts cold-start latency by up to \textbf{19.39$\times$}, enabling fast, cost-efficient scaling.

% A unified scaling controller dynamically adjusts resources in response to load and utilization across both colocated and disaggregated deployments, while also supporting seamless prefill–decode role transitions in P/D mode. Device-to-device (D2D) weight transfer further reduces cold-start latency by up to \textbf{19.39$\times$}, enabling fast and cost-efficient scaling in clusters.

\item \textbf{Operational Impact:} Evaluation under different workloads shows up to \textbf{4.44$\times$} higher SLO attainment, \textbf{65.82\%} lower request latency, and substantial operational cost savings compared to state-of-the-art baselines, demonstrating the effectiveness of our approach for real-world LLM serving.
\end{itemize}

%-------------------------------------------------------------------------------

%-------------------------------------------------------------------------------
% \input{osdi/motivation}
\section{Background and Motivation}
\label{sec:background}

\subsection{Large Language Model Serving}
The growing adoption of LLMs in interactive and high-throughput applications, such as chatbots, code assistants, and real-time translation services, places stringent demands on cloud serving infrastructure.

Most modern LLMs rely on the Transformer architecture and generate tokens in an \textit{auto-regressive} fashion. Inference with LLMs typically involves two computationally distinct phases: \textit{Prefill} and \textit{Decode}. 
\textbf{Prefill Phase}: Processes the initial prompt using full-sequence attention to generate the first token. This phase is compute-intensive and determines the TTFT. \textbf{Decode Phase}: The decode phase generates subsequent tokens incrementally, leveraging cached KV states and the latency of each token generation iteration determines the TPOT. This phase is memory-bandwidth bound which results in low arithmetic intensity that under-utilizes GPU compute resources.

LLM serving generally adopts either colocated or disaggregated deployments. Colocated mode \cite{kwon2023efficient} executes prefill and decode on the same devices. While this approach simplifies deployment and can reduce the number of devices required, it introduces the challenge that prefill-heavy requests can repeatedly interrupt ongoing decoding, causing TPOT to exceed its SLO. Conversely, disaggregated mode \cite{zhong2024distserve} isolates these phases to strengthen per-stage SLO control and scaling flexibility, but requires deploying model replicas across more devices. In workloads where either prefill or decode is lightly exercised, this separation can lead to underutilized resources. Disaggregation also introduces KV cache transfer overheads that can violate strict latency SLOs, so the optimal choice depends on specific TTFT and TPOT targets \cite{WangZuoChenLiangYuYang2025}. Overall, neither mode universally dominates, so effective serving systems must support both architectures to accommodate diverse production environments. It is important to note that this work does not aim to decide when colocated or disaggregated deployments should be used, nor does it address dynamic switching between the two. Instead, our goal is to ensure that multi-SLO scheduling and scaling operate effectively under both architectures, regardless of which deployment mode a production system adopts.

% \textcolor{red}{Niloofra; we have an explanation for the deployment mode in Introduction section, maybe double check, paragraph 4 Introduction.} Done edited in introduction

\begin{figure}[t]
    \centering
  \includegraphics[width=.66\columnwidth]{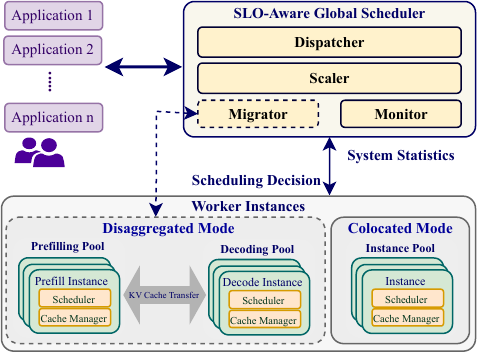}
  \caption{\textbf{Overview of the \paper{} architecture.} 
    The dispatcher, migrator, monitor, and scaler collectively enable proactive, SLO-aware scheduling and scaling. The system supports both disaggregated and colocated modes, and continuously aggregates system and SLO feedback for closed-loop adaptation under dynamic load.
    % {\color{red} sorry for lots of comments but small things make lots of difference. 1. yellow Boxes in instances are not aligned and do not have same size, 2. Cache Manager is better instead of cache engine, 3. green boxes in disaggregated P are not same as D. 4. Also in Colocated boxes are not good stacked on each other.  5. White boxes are not aligned between Disaggregated and collocated (grey boxes heights are not same.) better to have dash line for disaggregated same as migrator. 6. width of dispatcher and scaler are not same. 7. start position of migrator is not same as dispatcher. 8. We do not need border for the image and also image is not aligned in the column that is maybe because of some border in image.  9. Maybe don't need to put application image in boxes as two of them are circle and one of them is square also they are not in same size. 10. User icon height should increase to be more realistic. 11. KV cache transfer grey arrow should be between two grey boxes or inside them in both now it is just inside prefil pool box.}
    }
    
    \label{fig:overview}
\end{figure}

\subsection{Multi-SLO Scheduling Challenges}

To scale with increasing load, LLM serving systems run across clusters of many instances. However, standard load balancers (e.g., Round-Robin, Least-Connections) are insufficient for SLO-aware serving because they ignore the state-dependent nature of LLM workloads and cannot prioritize requests by remaining SLO budget. Dedicating instances to different request types or latency classes avoids interference but introduces new issues:

(1) resource under-utilization, as some instances idle while others queue; (2) operational inflexibility, requiring reconfiguration for new latency classes; and (3) higher provisioning cost due to per-class over-provisioning. Moreover, static separation fails to address the core issue: request urgency changes over time within each instance, requiring continuous scheduling decisions.

% (1) resource under-utilization, as some instances may remain lightly loaded while others accumulate long queues; (2) operational inflexibility, the deployment must be reconfigured when new latency classes are introduced; and (3) increased provisioning cost, since maintaining separate capacity for every request class results in over-provisioning during low-demand periods. Furthermore, this static separation does not solve the core problem: within a single instance, request urgency changes over time, and the scheduler must continuously decide. 

A second challenge arises from the nature of batching in modern serving systems. While frameworks like vLLM \cite{kwon2023efficient} and TensorRT-LLM \cite{tensorrt} adopt continuous batching to maximize the throughput, their perspective is strictly local. The scheduler within each instance treats the stream of incoming requests as fixed; it can determine how to batch what it receives, but not which requests it receives in the first place. Consequently, these systems lack the global context required to navigate trade-offs across multiple instances. For example, a latency-sensitive request may be queued behind long-context requests on one instance, even though another instance in the cluster has available compute capacity. 
Without a global scheduler that understands the current queue depth, memory fragmentation, and request types across all instances, the cluster suffers from Head-of-Line (HOL) blocking, where high-priority requests get stuck behind long-running ones on a selected instance.

\subsection{Elastic Scaling Challenges} 
\label{sec:elastic-challenge}
LLM workloads vary widely in request rate, length, priority, and latency SLOs. In large cloud deployments such as Huawei Cloud, clusters must handle highly dynamic traffic, making multi-instance scheduling even harder. Beyond these scheduling challenges, large-scale deployments must also contend with elastic scaling, which introduces a separate class of operational difficulties.

\textbf{Cold-Start Penalty:} Scaling is not instantaneous. New instances incur significant delays while cold-start overhead occurs when newly provisioned instances instantiate the inference engine and load model weights from storage.

\textbf{Cost vs. SLO Attainment Tradeoff:}
Scaling must balance the risk of SLO violations against the cost of idle resources, creating a core optimization challenge: providing enough capacity to meet deadlines without over-provisioning. Although this resembles a constrained scheduling and allocation problem, full end-to-end optimization is impractical due to three compounding factors. 

First, decisions are event-driven, triggered by requests for arrivals, completions, or the completion of an instance’s iteration, rather than occurring on a fixed rolling horizon \cite{vanHentenryck1996rolling}, resulting in an irregular decision timeline. 
Second, future workloads are uncertain, as request lengths are unknown and arrivals are highly stochastic.
Third, batching forms a circular dependency: given the current queue delays and SLO requirements, the scheduler chooses a batch size; this batch size determines the per-iteration latency, which then feeds back into future queue delays. This closed feedback loop has no simple analytic form. A robust system must jointly address \textit{micro-level scheduling} (routing requests to the best instance) with \textit{macro-level scaling} (detecting when current instances will no longer suffice).

Existing LLM serving systems do not fully address the operational complexities of large-scale cloud environments. One-shot dispatching and static scaling fail to handle dynamic queues, multi-stage workloads, and heterogeneous SLOs, leading to unpredictable latency, inefficient utilization, and higher costs. These gaps motivate \paper{}, a unified LLM serving system for Huawei Cloud that integrates multi-SLO-aware scheduling and elastic scaling for predictable, cost-efficient performance at production scale.

% Existing LLM serving systems, do not fully address the operational complexities observed in large-scale cloud environments. One-shot dispatching and static scaling strategies fail to account for dynamic queuing, multi-stage workloads, and heterogeneous SLOs, resulting in unpredictable latency, inefficient utilization, and increased operational costs. These gaps motivate the design of \textbf{\paper{}}, a unified LLM serving system for Huawei Cloud engineered to integrate multi-SLO-aware scheduling and elastic scaling for predictable performance and cost efficiency in production-scale deployments.

\begin{figure*}[htbp!]
    \centering
    \includegraphics[width=\linewidth]{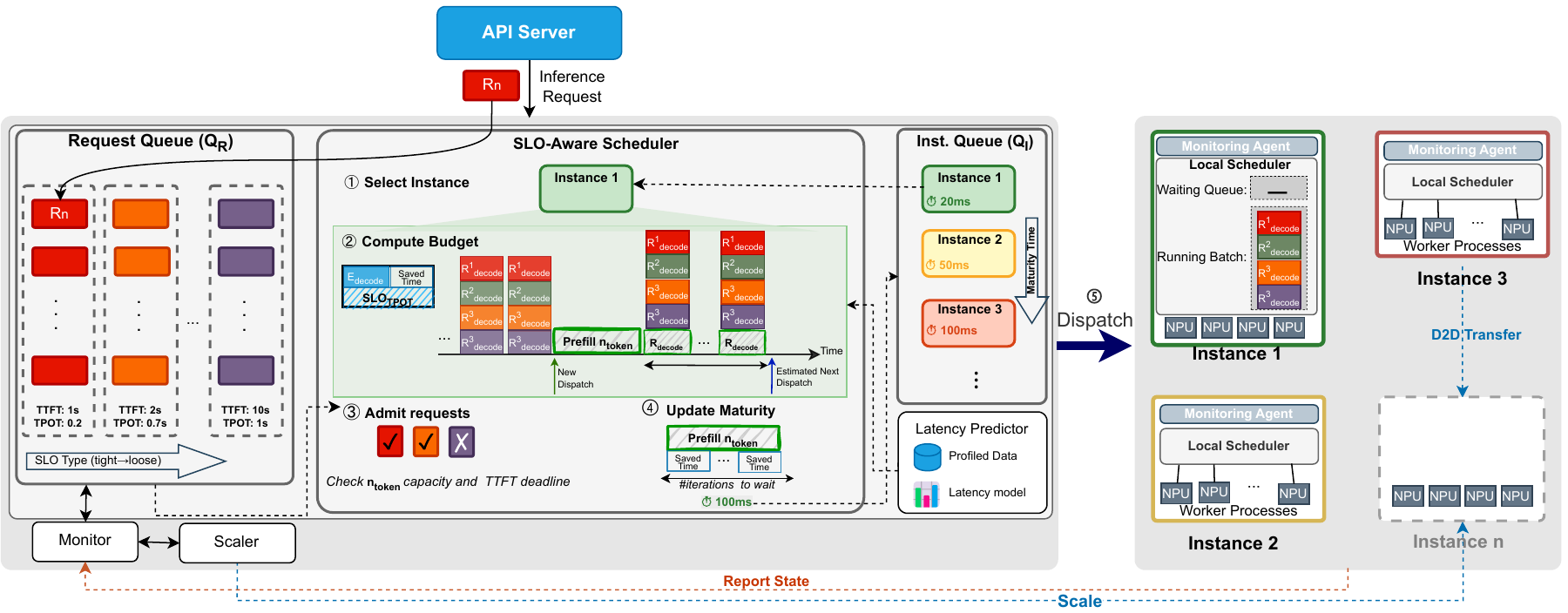}
    \caption{\textbf{Execution flow of \paper{}.} 
    A client request enters through the API Server into \textit{Dispatcher}'s \textit{Request Queue}. The \textit{SLO-Aware Scheduler} of \textit{Dispatcher}: (1) selects an instance from \textit{Instance Queue}, (2) computes the budget using the \textit{Latency Predictor} and system status, and (3) admits requests from \textit{Request Queue}. Then (4) updates the maturity time of selected instance, and (5) dispatches admitted requests to the selected instance.
    % \textcolor{red}{Niloofar: In inst.Queue, instances 1, 3, 4, etc. need to be corrected. Also, if we want to stay consistent with Figure 1, it might be good to include the following parts that are  missing in Figure 1: the Dispatcher’s Queue Manager and the latency predictor. They exist here, not in figure 1}
    }
    \label{fig:hyperflex}
        %  \cnumGreen{6}
\end{figure*}

\section{System Design}

% \textcolor{green}{--------------}
% \textcolor{green}{beginning of re-written section}
% \textcolor{green}{--------------}

\subsection{Problem Definition}
\label{sec:problemformulation}

{
% In this section, we first formalize the multi-SLO serving scenario and the SLO-aware scheduling problem. 
% % A small toy example then illustrates why simple scheduling strategies fail to balance cost and SLO attainment. 
% Because the scheduling dynamics are highly event-driven and difficult to capture in a clean mathematical program, 
% % we instead identify a set of core design principles that characterize the constraints any practical solution must satisfy. These principles guide the algorithms and mechanisms introduced in the following sections.
% we then identify core principles that guide the system and algorithm designs.
The serving system (Fig.~\ref{fig:overview}) receives a continuous stream of requests $R = \{ r_1, r_2, \dots, r_m, \dots \}$. 
Each request $r_m$ arrives at a certain timestamp and specifies its prompt, maximum output length, and heterogeneous SLOs, including TTFT and TPOT. 
Meanwhile, the system should support elastic scaling, adding instances on demand up to a maximum of $N$, constrained by available resources. 

At every event boundary, either a new request arrival or the completion of a prefill or decode iteration by any instance, the \textbf{Global Scheduler} should make two types of decisions:
(i) \emph{dispatching}: whether and how to assign pending requests to instances, and 
(ii) \emph{scaling}: whether to scale out or scale in, and which instances to deactivate when scaling in.
These decisions must balance two conflicting objectives: minimizing infrastructure cost while maximizing user satisfaction.
Specifically, infrastructure cost is given by 
\begin{equation}
\text{Cost} = \sum_{n=1}^N (\text{active time of instance } n)\cdot \text{UnitCost},
\end{equation}
where $\text{UnitCost}$ denotes the rental price per-second for a Neural Processing Unit (NPU)~\cite{liao2021ascend} instance.
Meanwhile, user satisfaction is measured by SLO attainment, defined as the fraction of requests that meet both the TTFT and TPOT requirements:
\begin{equation}
\text{Attainment} = 
\frac{1}{M} 
\sum_{m=1}^M 
\mathbb{I}_{\text{TTFT}}(r_m)\,
\mathbb{I}_{\text{TPOT}}(r_m).
\end{equation}

% \color{green} Xinglu: Should we consider skip this E2E metric here, and introduce it on demand in the experiment metrics section? }

Apart from SLO attainment, end-to-end (E2E) completion time for a request $r_m$ provides a complementary view of user-perceived latency, defined as 
${\text{E2E}}(r_m) = T_{\text{complete}}(r_m) - T_{\text{arrival}}(r_m).$
% \textcolor{cyan}{Since the overall objective of the system is to (1) Reduce NPU
% usage cost and (2) Maximize
% user satisfaction, I think it is important to include how we measure those in the problem formulation.}
% \textcolor{green}{Sounds good, let keep previous sentence. }

% As discussed in Section \ref{sec:elastic-challenge}, cost and satisfaction metrics reveal an inherent tradeoff. An effective scheduler must navigate this tension by admitting requests only when their SLOs can be met while avoiding unnecessary instance activity.
% Although this problem resembles a constrained scheduling and allocation task, a full end-to-end optimization is not tractable in practice. Several factors drive this complexity. 
% First, decisions are event-driven---triggered by request arrivals, completions, or the completion of an instance’s iteration---rather than occurring on a fixed rolling horizon \cite{vanHentenryck1996rolling}, resulting in an irregular decision timeline. 
% Second, it is difficult to accommodate future uncertainty, as output lengths are unknown at arrival and workload arrivals follow highly stochastic patterns.
% Third, batching forms a circular dependency: given the current queue delays and SLO requirements, the scheduler chooses a batch size; this batch size determines the per-iteration latency, which then feeds back into future queue delays. This closed feedback loop has no simple analytic form.

As discussed in Section \ref{sec:elastic-challenge}, lowering the cost and increasing the attainment of SLO has some challenges. These challenges make an explicit formulation infeasible. We therefore distill a set of core guiding principles to drive both the algorithmic and system design:
(1) At \textit{micro-level scheduling}, the system prioritizes protecting in-flight SLOs, ensuring that admitted tasks can continue progressing toward their SLO requirements before any additional load is accepted.
Meanwhile, new requests are admitted greedily, whenever it is safe, to maximize hardware utilization and thereby reduce cost.
(2) At \textit{macro-level scaling}, Elastic scaling serves as a fallback when no safe placement exists, enabling the system to adjust resources dynamically while preserving SLO compliance.
These principles are instantiated naturally in both colocated and disaggregated deployment modes, providing a practical framework for building an SLO-aware serving system that balances user satisfaction and cost efficiency in dynamic workloads.

%-------------------------------------------------------------------------------

\subsection{\paper{} Overview}
\label{sec:overalldesign}

In this section, we present the overall system design and framework of \paper{}. 

% {\color{green} xinglu: There is a rephrase of principles in previous section, maybe we can commented out following one.}
% \textbf{Principles.} We present our design based on two guiding principles:
% \begin{enumerate}
%     \item Locally, we ensure that each instance adheres to SLOs for its requests, maximizing the number of requests it can process without violating these constraints.
%     \item Globally, a scaler makes scaling decisions based on the current system state, scaling out to handle urgent requests and scaling in during under-utilization.
% \end{enumerate}

\textbf{Architecture.} As shown in Figure~\ref{fig:overview}, \paper{} comprises four main components: the \textit{Dispatcher}, \textit{Scaler}, \textit{Monitor}, and \textit{Migrator}. 
The Dispatcher manages request scheduling, while the Scaler dynamically adjusts the number of active instances and enables efficient weight loading during scale operations. 
Both rely on system statistics from the Monitor to make informed decisions. 
The Migrator coordinates task dispatching in the decode phase, links prefill and decode instances, and manages KV-cache transfers to ensure smooth, efficient execution.

\textbf{Workflow.} Figure~\ref{fig:hyperflex} illustrates \paper{}’s execution flow. Multiple clients submit LLM inference requests with heterogeneous latency targets, including TTFT and TPOT. Requests enter via an API server and are managed by the centralized Multi-SLO-Aware Dispatcher (\S\ref{sec:dispatching}), which classifies requests by SLO and enqueues them in a priority queue. The dispatcher selects instances based on predicted availability (\textit{maturity time}) and dispatches requests using system status, token budgets, and predicted latencies, ensuring SLOs are not violated for new or ongoing tasks.

Execution depends on deployment mode. In colocated setups, both TTFT and TPOT are considered; in disaggregated P/D mode, the dispatcher considers only TTFT and assigns requests to Prefill instances. The \textit{Migrator} then handles TPOT, migrating requests from Prefill to selected Decode instances as shown in Figure~\ref{fig:hyperflex}.

Resource provisioning is handled by \textit{Scaler} (\S\ref{sec:scaling}) for both colocated and disaggregated deployments. It adjusts instance counts dynamically based on utilization and queue metrics and supports seamless role transitions between prefill and decode instances.

% Resource provisioning is managed by a Scaling Controller (\S\ref{sec:scaling}), which supports both disaggregated and colocated deployments. It dynamically adjusts instance counts based on real-time metrics such as utilization and queue states. The Scaling Controller also enables seamless role transitions between prefill and decode instances.

% To further improve scalability, \paper{} integrates a fast scaling mechanism. Instead of loading model weights from storage, weights are transferred directly from an already running instance via a high-bandwidth D2D link, significantly reducing cold-start latency.

\textit{Monitor} subsystem underpins these mechanisms by continuously collecting statistics on instance utilization, request latencies, queue states, and instance roles. This data feeds into both scheduling and scaling decisions.

In summary, \paper{} aims to meet diverse latency objectives while using resources efficiently. The following section describes how these mechanisms are implemented in practice.

% In summary, \paper{} unifies multi-SLO-aware scheduling, disaggregated prefill/decode execution, and fast, role-adaptive scaling into a single system. It ensures efficient resource use while meeting diverse latency objectives. In the following section, we elaborate on how these mechanisms are realized in practice. 

% {\color{red} morgan: this overview is really long. I think we can condense here.... \paper{} unifies multi-SLO-aware scheduling, elastic scaling, and disaggregated prefill/decode execution into a single framework. Requests are assigned to instances by the Dispatcher (detailed in \S\ref{sec:dispatching}) based on predicted latencies and token budgets, while the Migrator coordinates KV-cache transfers in disaggregated mode. The Scaling Controller (detailed in \S\ref{sec:scaling}) dynamically adjusts instance counts and roles, leveraging fast D2D weight transfer to reduce cold-start latency. The Monitor continuously collects statistics on utilization, latencies, and queue states to inform both scheduling and scaling decisions. Priority assignment and SLO mapping are described in \S\ref{sec:pmap}. }

\subsection{Detailed System Design}
\subsubsection{Dispatching Mechanism and Policy}
\label{sec:dispatching}

Our SLO-aware dispatcher (i) schedules new requests promptly so latency-critical work begins without unnecessary delay, while (ii) ensuring ongoing executions meet their SLOs. However, scheduling a new request can interfere with in-progress work, increasing latency and risking SLO violations. To manage this, we introduce \emph{maturity time}, the earliest an instance can take additional load without affecting ongoing requests. By tracking maturity times centrally, the dispatcher enables low-overhead, responsive global scheduling that protects existing workloads.

% Our dispatching mechanism is SLO-aware and aims to (i) admit new requests promptly so latency-critical work begins without unnecessary delay, while (ii) ensuring ongoing executions continue to meet their SLOs. The difficulty is that admitting a new request can interfere with work already in progress by consuming compute and memory bandwidth, thereby increasing latency and risking deadline violations. To manage this interference, we introduce the notion of \emph{maturity time}, which denotes the earliest point at which an instance can accept additional load without risking the SLOs of in-flight requests. By maintaining and updating maturity times centrally, the dispatcher enables low-overhead global scheduling that is both responsive and protective of existing workloads.

Since colocated and disaggregated deployments exhibit different interference behaviors, we apply separate control policies for each mode.

\paragraph{Colocated Mode.} 
In colocated deployments, prefill and decode share the same device. New requests trigger prefill computations that can preempt ongoing decoding, increasing latency and risking TPOT violations. The dispatcher’s main role is to \emph{protect ongoing decode workloads} while scheduling urgent new requests before violating their TTFTs.

% In colocated deployments, prefill and decode run on the same device. A newly arriving request immediately triggers a prefill computation, which can preempt and slow ongoing decoding, increasing per-token latency and potentially violating TPOT targets. The dispatcher’s primary responsibility in this mode is therefore to \emph{protect ongoing decode workloads} while still admitting urgent new arrivals as soon as it becomes safe.

The dispatcher centrally tracks all instances and controls when each can accept new prefills. For each instance, it computes a maturity time, the earliest point at which new requests can be admitted without violating ongoing decode SLOs. Before maturity, the instance is protected; afterward, it can take new assignments. We also use \emph{token budget} to represent the maximum number of tokens can be dispatched to a instance without violating ongoing decode TPOT limits. Next, we will first give a high-level overview of the dispatching steps, then explain how token budget and maturity time is computed.

% The dispatcher maintains centralized visibility over all instances and regulates when each may accept additional prefills. For each instance, it computes a \emph{maturity time}: the earliest future point at which admitting new requests will not cause current decode jobs to violate their SLOs. Before maturity the instance is protected; after maturity it is again eligible for new assignments.

% \sout{Maturity time is determined by estimating how many new requests can be scheduled to an instance. We use a \emph{token budget}, representing the additional prefill-induced delay an instance can absorb without violating ongoing decode TPOT limits. This budget reflects slack from prior progress and the ability to spread small interruptions over the remaining decoding window before maturity point.}

% Determining maturity time requires estimating how many new requests an instance can safely admit. We capture this using a \emph{token budget}, which represents the maximum additional prefill-induced delay that can be absorbed without pushing ongoing decode requests beyond their TPOT limits. This budget reflects the amount of slack created by prior progress and the ability to amortize small interruptions over the future uninterrupted decoding window before the maturity point.

% \sout{With the token budget available, the dispatcher repeatedly executes the following steps:}
% Once the dispatcher knows the token budget, it can decide how many, and which, new requests can be admitted without violating SLO constraints. 
% At a high level, the dispatcher repeatedly performs four decisions:
Dispaching steps are as follows: 
(1) select the instance whose maturity time arrives first;
(2) compute its safe token budget;
(3) scan the global request queue to identify the most urgent requests that fit within this budget; and
(4) dispatch them, then update the instance’s future maturity time based on the newly predicted workload.
This process ensures that decode jobs remain protected, while unused capacity is quickly reclaimed to serve urgent incoming requests. The detailed realization in colocated mode is shown in Algorithm 1 in the supplementary material.

% {\color{green}
We now explain how the token budget is computed as shown in Step 1 of Figure~\ref{fig:hyperflex}. Firstly, we introduce \textit{slack time}, as the saved time of each decoding token compared to the TPOT requirement. With decoding iterations, the slack time accumulates, creating a `safe' time window for new Prefill batch to jump in, without violating the ongoing decoding TPOT. This is the core concept of the algorithm. 

Next, lets first assume the dispatcher has decided to dispatch $n_{\text{token}}$ tokens in the new prefill batch to the instance (later we will explain how $n_{\text{token}}$ is computed). Then the prefill latency of the new batch $E_p$, and the decoding latency of following decoding iterations $E_d$ can be estimated following \cite{zhong2024distserve}. Referring to the core concept, $E_p$ is actually the `safe' time window for the new prefill batch, and $\text{TPOT} - E_d$ will be the slack time of each following decoding iteration. Then $k = E_p / (\text{TPOT} - E_d)$ decoding iterations will be needed to accumulate enough slack time to achieve the `safe' time window. Before $k$ decoding iterations, the instance cannot admit new requests. The \textbf{maturity time} is the point at which the instance becomes eligible to admit new requests again. It is the sum of the prefill latency and the slack-accumulating decoding iterations:
\[\text{Maturity Time} \coloneqq E_p + k \cdot E_d.\] Note that, for simplicity, we ignored any past-accumulated slack time before current dispatching. As our scaling algorithm will guarantee system always runs in nearly-full capacity, the past-accumulated slack should be minimal.

Now, we revisit how to compute $n_{\text{token}}$, which we define as \textit{token budget} for the instance. Inherently, the larger $n_{\text{token}}$, the longer maturity time will be, and it will delay the instance to accept new requests, consequently, hinders future request's TTFT. So, we should limit $n_{\text{token}}$. We adopt a worst-case guarantee methodology. The worst case for future request is that a request arrives right after the current dispatching, then the estimated time for its actual TTFT will be: $\text{Maturity Time} + E_p = 2E_p+k \cdot E_d$. Considering the TTFT constraint, we get
\[
2E_p + k \cdot E_d \le \text{TTFT}.
\]

As shown in the Appendix~A.2 of DistServe \cite{zhong2024distserve}, prefill latency can be estimated using a quadratic performance model. However, based on both our NPU profiling and their released code, the quadratic term is negligibly small in practice. We therefore adopt a linear model $E_p = a + b\,n_{\text{token}}$. And as $E_d$ grows slowly with new request added, we can ignore the effect of new $n_{\text{token}}$ and use current running batch to estimate a constant $E_d$. Substituting them into the TTFT inequality and simplifying yields the token-budget constraint in Eq.~\ref{ntoken}:

\begin{equation} \label{ntoken}
n_{\text{token}} \le 
% \frac{\text{TTFT}\cdot\text{TPOT} 
%       - \text{TTFT}\cdot E_d 
%       - a\cdot\text{TPOT}}
%      {b\cdot\text{TPOT}}.
\frac{(\text{TPOT} - \text{E}_d)\cdot \text{TTFT}}{(2\text{TPOT} - \text{E}_d) \cdot b} \;-\; \frac{a}{b}.
\end{equation}

Thus, we get an upper bound for $n_{\text{token}}$, and the dispatcher can dispatch accordingly.

\paragraph{Disaggregated Mode.}
\label{sec:pddisagg}
 
In the PD-disaggregated deployment mode, prefill and decode run on separate instance pools, so the four-step dispatching logic in colocated mode is instantiated twice: once by the \textit{Dispatcher} for the P stage and once by the \textit{Migrator} for the D stage.
Both components follow the same four-step structure, with only Step~2 and Step~4 differing from the colocated mode due to a different interference model. We therefore focus on explaining these two steps for each stage.

\emph{P-stage Dispatcher.} 
Prefill batches are non-interruptible under static batching. Consequently, the \textit{Dispatcher} uses a simplified maturity definition: an instance becomes mature only when its current prefill batch completes. The different steps are:
\begin{itemize}
   \item \emph{Step~2 (token budget).} Starting from an empty instance, its safe token budget is determined solely by the strictest TTFT in the priority-ordered request queue. The intuition is that lower-priority requests have looser TTFT requirements, so satisfying the strictest TTFT automatically ensures that remaining requests meet their TTFT constraints. 
   \item \emph{Step~4 (maturity time).} After dispatching a new batch, the instance’s next maturity time is set to the estimated completion time of this prefill batch, since it cannot be interrupted mid-way.
\end{itemize}

\emph{D-stage Migrator.} 
Under continuous batching, decode progress is iterative, and each iteration boundary provides a natural decision point, enabling fine-grained control. 
The \emph{Migrator} makes the dispatching decision. 
The different steps are:
\begin{itemize}
    \item \emph{Step~2 (token budget).} The safe token budget depends primarily on the state of the decoding instance. When the instance already has an active decode batch, its effective TPOT constraint is determined by the strictest TPOT among the in-flight requests, and the budget is computed to ensure that admitting additional tokens will not violate this constraint. If the instance is temporarily idle, it instead uses the strictest TPOT from the priority-ordered queue of prefilled requests to determine its token budget.
    \item \emph{Step~4 (maturity time).} After dispatching, maturity is updated to the end of the next decoding iteration, enabling a decision at every iteration boundary.
\end{itemize}

Simple strategies such as round-robin or dispatching to the least-loaded instance do not consider instance states or request-level SLO requirements, and therefore may admit requests even when the resulting latency cannot meet SLO constraints. In contrast, our dispatching logic is SLO-aware and avoids these limitations.

\subsubsection{Priority-Based SLO Mapping}
\label{sec:pmap}

\begin{figure}[tb]
    \centering
    \includegraphics[width=\linewidth]{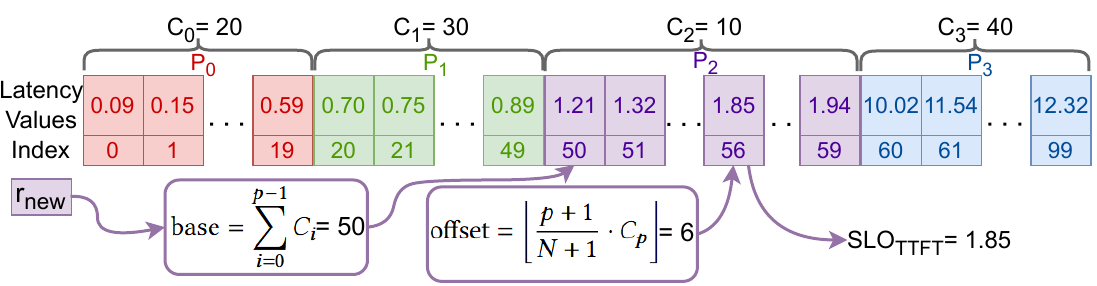}
    % \Description{Example schematic of priority-based SLO mapping for the TTFT (time-to-first-token) SLO. The diagram shows several priority levels (denoted by the different colours, with corresponding TTFT targets and their index. This is the recent performance data that has been placed into the sliding windows, illustrating how higher-priority requests have stricter TTFT targets, while lower-priority requests have more relaxed targets. On the bottom we can see how the TTFT of a new priority 2 request would be assigned. This mapping demonstrates how the system adjusts latency objectives based on request priority.}
    \caption{\textbf{Priority-based SLO mapping example.} $\text{SLO}_{\text{TTFT}}$ is determined from recent performance data.}
    \label{fig:pmap-example}
\end{figure}

We design a priority-based SLO mapping scheme to handle cases where users cannot specify explicit metrics such as TTFT or TPOT. In real deployments, customers often lack precise latency targets, but they can usually express the relative importance of their applications. Our scheme leverages these priorities to derive latency SLOs dynamically, adapting them over time as the system operates. This is illustrated in Figure~\ref{fig:pmap-example} and explained below with five steps.

\textit{(1) Priority as input.} Each request is assigned a priority $p$, with $0$ being the highest and $N-1$ the lowest, with $N$ representing the total number of priority levels.
\textit{(2) Sliding-window history.} The system maintains two sliding windows of size $W$ for recent TTFT and TPOT values. Each new result is inserted while the oldest entry is dropped, ensuring SLOs are based on up-to-date conditions. 
\textit{(3) Mapping priority to latency.} For each new request ($r_{\text{new}}$), the system determines its SLO latency within the window using an indexing scheme. The base index accounts for all higher-priority requests already placed in the window, while the offset determines the request’s position within its own priority class. This mapping biases higher priorities toward lower-latency quantiles and lower priorities toward higher-latency quantiles. Formally:  
\begin{equation}
\text{base}   = \sum_{i=0}^{p-1} C_i, \quad
\text{offset} = \left\lfloor \frac{p+1}{N+1} \cdot C_p \right\rfloor
\end{equation}

where $C_i$ is the count of completed requests with priority $i$ in the window. The sum ($i_s=\text{base} + \text{offset}$) indexes the latency used as the provisional SLO for $r_{\text{new}}$. 
\textit{(4) Correcting for queuing effects.} To prevent temporary queue spikes from inflating SLOs, we subtract the extra queuing delay between the chosen reference request and the most recent request of the same priority. This correction keeps SLOs aligned with real-time system behavior.
\textit{(5) Ensuring stability.} Derived SLOs are bounded within pre-defined ranges per priority. A contention-aware rule further stabilizes outcomes: If higher-priority requests are pending, the system enforces the strict lower bound. Otherwise, the bound is relaxed so lower-priority requests may complete sooner without undermining priority ordering.

Finally, the adjusted TTFT and TPOT values are delivered to the SLO-aware scheduler for integration into dispatching decisions. The detailed algorithm is provided in Algorithm 3 of the supplementary material.

% \renewcommand{\algorithmcfname}{Algorithm}
% \SetKw{KwAttribute}{Attribute}

% \begin{algorithm}[t] 
% \caption{Priority-Based SLO Assignment} \label{algo:priority-slo-mapping} 

% \KwIn{Request $r_{new}$ with priority $p \in [0, N{-}1]$}
% \KwOut{Assigned SLOs for request $r_{new}$}

% \KwAttribute{\texttt{ttft\_records} and \texttt{tpot\_records}: }{ Sorted sliding windows of recent request SLOs (size $W$)}

% \KwAttribute{$C_0, C_1, ..., C_{N-1}$: }{ Counts of requests per priority in sliding window}

% \KwAttribute{\texttt{last\_queue\_time}: }{ Last observed queue time list for each priority}

% \tcp{Retrieve TTFT and TPOT from sliding windows}
% $idx \gets get\_priority\_index(p, C[0:N-1])$\;
% $(ttft, q\_time) \gets \texttt{ttft\_records}[idx]$\;
% $tpot \gets \texttt{tpot\_records}[idx]$\;
% \tcp{Adjust TTFT to reduce impact of queue time spikes}
% $\Delta_{queue\_time} \gets q\_time - \texttt{last\_queue\_time}[p]$\;

% $ttft \gets ttft - \Delta_{queue\_time}$\;
% \texttt{last\_queue\_time}[p] $\gets q\_time$\;

% \tcp{Retrieve and clamp TTFT and TPOT} 
% $ttft \gets clamp(ttft, \texttt{min\_ttft}(p), \texttt{max\_ttft}(p))$\;
% $tpot \gets clamp(tpot, \texttt{min\_tpot}(p), \texttt{max\_tpot}(p))$\;

% \Return $(ttft, tpot)$\;

% \end{algorithm}

\subsubsection{Scaling Mechanism and Policy}
\label{sec:scaling}

% {\color{green} From Parham's advice:
% It is not clear how our scaling is accounting for different SLOs by considering all these metrics together. Also is scaling triggered if they are "above" threshold? for example if system processing rate is above threshold it is good signal, why is it considered as a signal of overload? 

% Maybe conciser this: 

% Dynamic scaling is essential for maintaining TTFT/TPOT attainment under fluctuating load. Insufficient resources lead to queueing delays and propagated SLO violations, whereas excessive provisioning increases cost. This section presents our design of \emph{algorithmic} and \emph{system-level} designs that enable responsive and cost-efficient scaling. The algorithm determines when and how to resize prefill and decode capacity, and the system mechanisms ensure that scaling decisions can be executed quickly and reliably.
Dynamic scaling is key to maintaining SLOs during load fluctuations. Too few resources cause queueing and SLO violations; too many increase cost. We present \emph{algorithmic design} and \emph{system-level techniques} that enable fast, cost-efficient scaling: the algorithm decides when and how to scale prefill and decode capacity, while the system executes these decisions efficiently and reliably. 
We focus on the algorithmic design here and leave the implementation details of the system techniques to Section~\ref{sec:impl}.

% \paragraph{Algorithmic Design}
% At a high level, the scaling logic follows Algorithm~\ref{algo:scaling}. 
The detailed pseudocode is provided in Algorithm 2 of the supplementary material; we present its decision logic here. 
Every~$\tau$ seconds, the \textit{Scaler} inspects the global request queue and all active instances to compute three indicators for scaling.
\begin{itemize}
    \item The first captures \textit{system load} over a \textit{long} (10\,s) horizon. 
    The Scaler maintains a fixed 10\,s window of enqueue and dequeue timestamps to derive smoothed arrival and completion rates, $\lambda_{\text{in}}$ and $\lambda_{\text{out}}$. 
    To avoid spurious actions during startup or with few events, rate estimates are suppressed until enough samples exist. 
    The load estimator is $f_1 = \frac{\lambda_{\text{out}}}{\lambda_{\text{in}}}$: $f_1<1$ indicates completions lag arrivals (pressure to scale out), and $f_1>1$ indicates excess capacity.
    
    \item Second, we use a \textit{SLO-aware load indicator} that captures \textit{short-term} fluctuations. 
    The Scaler computes the normalized waiting time of each queued request relative to its SLO, $f_2 = \frac{1}{N} \sum_{r \in \text{queue}} \frac{\text{waiting\_time}(r)}{\text{SLO}_r}$, 
    where $N$ is the number of queued requests. 
    Higher $f_2$ indicates more urgent requests and greater risk of SLO violations.

    \item The third indicator captures each \textit{instance's memory load}. 
    For instance~$n$, the Scaler maintains a coarse utilization estimate based on KV-cache usage, $f_3 = \frac{\text{used\_blocks}_n}{\text{total\_blocks}_n}$, where $\text{used\_blocks}_n$ includes blocks held by in-flight requests and those reserved for pending requests in the instance's local queue.
 
\end{itemize}

These three indicators drive a threshold-based scaling policy. 
$f_1$ is the primary rate signal: $f_1 < \epsilon_{\text{out}}$ triggers scaling out, while $f_1 > \epsilon_{\text{in}}$ and multiple instances allow scaling in by removing the instance with smallest $f_3$. 
The SLO-aware signal $f_2$ reacts quickly to bursts: $f_2 > \epsilon_{\text{wait}}$ triggers proactive scale-out even if $f_1$ is moderate. 
By default, we use a 10\,s smoothing window, $\epsilon_{\text{out}}=0.7$, $\epsilon_{\text{in}}=1.1$, and $\epsilon_{\text{wait}}=1/4$ of the SLO target.

% These three indicators drive a threshold-based scaling policy. $f_1$ serves as the primary rate signal: when $f_1 < \epsilon_{\text{out}}$, the system interprets sustained overload and activates an additional instance. Conversely, when $f_1 > \epsilon_{\text{in}}$, and more than one instance is active, the Scaler scales in by selecting the instance with the smallest $f_3$, as it carries the least KV-cache commitment. 
% Meanwhile, the SLO-aware signal $f_2$ provides fast reaction to bursts: if $f_2 > \epsilon_{\text{wait}}$, the system proactively scales out even when $f_1$ remains moderate, in order to satisfy user SLOs under short-term fluctuations. 
% For these three hyper-parameters, by default, we use a smoothing window of $W=10$\,s, a rate-based threshold pair of $\epsilon_{\text{out}}=0.7$ and $\epsilon_{\text{in}}=1.1$, and an SLO safeguard threshold of $\epsilon_{\text{wait}}=\frac{1}{4}$ (i.e., $\frac{1}{4}$ of the SLO target).

In PD-disaggregated mode, prefill and decode stages have different workloads, so the Scaler evaluates each stage independently. 
The two instance pools may diverge—for example, prefill may scale in while decode scales out—unlike in the colocated case. 
To avoid unnecessary churn from terminating and relaunching instances, the Scaler can make a \emph{role switching decision}, e.g., reassigning a prefill instance's NPU resources to the decode pool.

\subsection{Implementation} \label{sec:impl}

\paper{} is implemented in Python and consists of approximately 3.5K lines of code (LoC) across different modules, including the dispatcher, scaler, migrator, monitor, and instance runtime, as shown in Figure \ref{fig:overview}. The serving backend is built on vLLM-Ascend \cite{vllm-ascend}, extended to support multi-SLO-aware scheduling, fast role transitions, and device-to-device data movement across Ascend 910B NPUs \cite{liao2021ascend}. The entrypoint exposes an OpenAI-compatible API \cite{openai2020api} for inference and management. Each module is implemented as an independent Ray \cite{moritz2018ray} actor, allowing failures, restarts, and instance elasticity to be handled transparently at runtime. All actors communicate with asynchronous RPCs, and the system uses a multi-process design that isolates CPU-bound control logic from NPU-bound execution.

The instance runtime is a modified vLLM-Ascend engine running as a Ray actor bound to a configurable number of NPU devices. Each instance maintains a lightweight execution loop for prefill and decode, exposes token-level progress metrics to the scheduler, and integrates a KV cache manager backed by LLM-DataDist \cite{hiascend2025llmdatadist}, an Ascend runtime component for cross-process memory mapping and KV data movement. \paper{} extends this mechanism with a metadata exchange protocol and per-request KV-transfer parameters, supporting efficient KV exchange in PD-disaggregated execution.

% Scaling and device provisioning are implemented by a two-stage weight transfer pipeline inside the Scaler. Warm-up instances initialize Python runtimes without loading weights; when scaling out, Scaler establishes temporary communication groups and triggers HCCL-based device-to-device weight transfer from existing instances. A failure-handling mechanism falls back to disk-based weight loading when necessary. This optimization significantly reduces cold-start latency and is fully integrated into the disaggregated prefill–decode architecture, enabling rapid role reassignment without restarting engines. 
% {\color{green} xinglu: i moved system techniques in previous scaling section to here. Maybe conciser replace with following paragraph:

We design the system for fast scale-out, allowing new instances to start with minimal cold-start latency. 
This is achieved through a two-stage weight transfer pipeline inside the Scaler.
In the first stage, each NPU server maintains a warm-up pool of lightweight instances that initialize the Python runtime but defer model weight loading.
When scaling out (the second stage), the Scaler establishes temporary communication groups and triggers HCCL-based device-to-device \texttt{send\_weight}/\texttt{receive\_weight} transfers from existing instances. 
Specifically, each instance embeds a \textit{WeightManager} that tracks device memory layout, enabling any running instance to serve as a source for weight transfer. 
Tensors are copied in a device-aligned manner, allowing the target instance to join the running pool immediately after transfer.
If a transfer fails, the system falls back to disk-based weight loading to ensure correctness.
This mechanism significantly reduces cold-start latency and supports rapid role reassignment in the disaggregated prefill–decode architecture.

In PD-disaggregated mode, we further accelerate KV-cache transfer through the following system design. 
The \textit{Migrator} manages the communication-group topology for KV-cache transfers between prefill and decode instances. 
When executing a \emph{role-switch decision}, the Scaler reassigns an instance's role between prefill and decode, and the Migrator proactively establishes the necessary link groups before the instance begins serving, enabling KV-cache movement to proceed without delay.
In contrast, prior systems such as Mooncake~\cite{qin2024mooncake} construct these links on demand, incurring additional latency.

\paper{} is currently undergoing production deployment within Huawei ModelArts~\cite{huaweimodelarts}. Our scheduler is integrated with the existing ModelArts runtime, which provides Kubernetes-based orchestration, pod lifecycle management, and system monitoring. To support this deployment, we implemented a 8.3K-LoC Go control-plane component that augments the ModelArts CRDs and coordinates vLLM-Ascend instances, exposing the scheduling decisions produced by our system. The underlying hardware and communication stack (Ascend NPUs, HCCL, LLM-DataDist) remain unchanged. The deployment is proceeding through staged evaluation, replacing the scheduling layer while preserving the rest of the production infrastructure.

\begin{table}[tbp]
\centering
\setlength{\tabcolsep}{2.2pt}
\begin{tabular}{@{}lccccc@{}}
\toprule
\textbf{Task} & \multicolumn{2}{c}{\textbf{SLO (s)}} & \multicolumn{2}{c}{\textbf{Lengths}}   \\ \cmidrule{2-3} \cmidrule{4-5}
& TTFT & TPOT & Input & Output \\
\midrule
\texttt{med\_qa} \cite{cmekg2023}        &   0.7    &   0.5    &    32.6 ± 10.3   &   38.9 ± 16.8     \\
\texttt{tldr\_c}$^\dagger$ \cite{loraland2024} &    1   &    0.7   &   44.4 ± 6.6    &  96.0 ± 35.0  \\
\texttt{tldr\_h}$^\dagger$ \cite{loraland2024} &   2    &   0.9    &   121.8 ± 35.0    &  13.6 ± 6.6  \\
\texttt{wikisql} \cite{loraland2024}            &    20   &   1    &  643.2 ± 337.0     &   27.8 ± 4.8   \\
\hline
\texttt{gsm8k} \cite{cobbe2021training}            &  0.7    &    0.2   &  51.4 ± 15.8     &   90.1 ± 26.7    \\
\texttt{sharegpt}\cite{sharegpt2023} & 2 & 0.5 &  259.2 ± 324.9
   & 207.8 ± 235.0  \\
\bottomrule
\end{tabular}
\setlength{\tabcolsep}{6pt} 

{\raggedright {$\dagger$ tldr\_c and tldr\_h correspond to the \texttt{tldr\_content\_gen} and \texttt{tldr\_headline\_gen} datasets respectively}\par}
% \caption{Summary statistics of the benchmark tasks. The first four tasks form the 4-task multi-SLO set, and the last two form the 2-task set. Input/output lengths are mean $\pm$ standard deviation over 300 requests.}
\caption{\textbf{Summary statistics of the benchmark tasks.} The first four tasks form the 4-task multi-SLO set, and the last two form the 2-task set. Input/output lengths are reported as mean $\pm$ standard deviation over 300 requests.}
\label{tab:task-stats}
\end{table}

\section{Evaluation}
\label{sec:experiments} 
% \textcolor{blue}{Parham: Added an intro and key questions of our evals.}
Our evaluation aims to systematically assess the performance and resource efficiency of \paper{} under realistic, multi-SLO distributed inference workloads. Specifically, we seek to answer three core quantitative questions: (1) Does \paper{} achieve superior SLO attainment compared to competitive baselines? (2) How does \paper{}'s scheduling impact overall end-to-end latency for a heterogeneous mix of tasks? (3) Can \paper{} efficiently manage the dynamic scaling of resources to minimize resource cost while maintaining high SLO attainment? The following subsections detail the experimental setup, serving models, and benchmark workloads before presenting the results for these key metrics.
Beyond core performance, we also investigate the practical robustness and adaptability of \paper{} to dynamic production conditions. Specifically, we evaluate whether the system can (1) adapt to dynamic traffic patterns, (2) remain stable even when operating in single-task regimes it was not optimized for, and (3) preserve performance when key control parameters, such as monitoring and scaling intervals, are varied.

%-------------------------------------------------------------------------------

%-------------------------------------------------------------------------------
\subsection{Experimental Setup} \label{sec:setup}
We evaluate \paper{} on realistic multi-SLO workloads across multiple models and hardware setups. All methods run on the same cluster with identical configurations, and each experiment is repeated three times; plots show the mean with standard deviation as error bars.

\textbf{Hardware.}
% \textcolor{blue}{Parham Added:}
The experimental testbed consists of two Huawei Atlas 800I compute nodes \cite{huaweiatlas}. Each node is equipped with eight Ascend 910B NPUs \cite{liao2021ascend}, with 64GB of HBM capacity per NPU. Intra-node communication is facilitated by direct HCCS links providing a peak bandwidth of 720 Gbps. For inter-node communication, the system utilizes a RoCE-enabled Ethernet fabric via 200GE QSFP ports. 
This configuration is representative of high-performance cluster in the production  distributed inference clusters.
% We evaluate \paper{} under production conditions, testing multiple models, realistic hardware constraints, and heterogeneous multi-SLO workloads. All methods are deployed on the same cluster with identical software, hardware, and workload configurations to ensure fair comparison. Each experiment is repeated three times; plots show the mean with standard deviation as error bars.

\textbf{Models.}
We evaluate \paper{} on three widely used open-source LLMs: Qwen7B, Qwen32B~\cite{bai2023qwen}, and Llama70B~\cite{touvron2023llama}, covering lightweight to large-scale deployments. Inference uses BF16 precision for efficiency and accuracy, with tensor parallelism~\cite{shoeybi2019megatron} configured as TP=1, 2, and 8, respectively.

% Inference is performed with FP16 or BF16 precision, the standard configuration for inference serving as it balances efficiency and accuracy. We employ tensor parallelism~\cite{shoeybi2019megatron} (TP) to distribute weights across devices, using TP=1 for Qwen7B, TP=2 for Qwen32B, and TP=8 for Llama70B.

\textbf{Benchmark Workloads and Datasets.}
Production LLM services typically host multiple tasks with distinct latency targets and
input/output characteristics. To capture this variability, we evaluate two multi-SLO task
sets observed on Huawei Cloud. They contain diverse tasks with heterogeneous SLO requirements, summarized in Table~\ref{tab:task-stats}. The first set includes the \texttt{medical\_qa} dataset~\cite{cmekg2023}, which consists of auto-generated answers to clinical medical questions; the \texttt{tldr\_content\_gen} and \texttt{tldr\_headline\_gen} datasets~\cite{loraland2024}, which focus on text generation by producing article content from headlines and headlines from article content, respectively; and the \texttt{wikisql} dataset~\cite{loraland2024, zhong1709seq2sql}, a semantic parsing benchmark mapping natural language questions to SQL queries over tables. The second set comprises two tasks, \texttt{gsm8k}~\cite{cobbe2021training} that contains grade school math word problems and \texttt{sharegpt}~\cite{sharegpt2023} that is a collection of conversational data, representing high-complexity reasoning and conversational AI workloads, respectively. For evaluation, we draw 300 samples per task. Requests for each task follow a Poisson inter-arrival time distribution, with each task contributing an equal number of samples. A fixed random seed is used to ensure reproducibility.

For priority-based SLO mapping (\paper{}-PM), we use the median SLO targets from Table~\ref{tab:task-stats} with a $\pm 25\%$ range to separate priorities while allowing dynamic assignment based on real-time workloads.

\begin{figure*}[tbp]
    \centering
    \includegraphics[width=\linewidth]{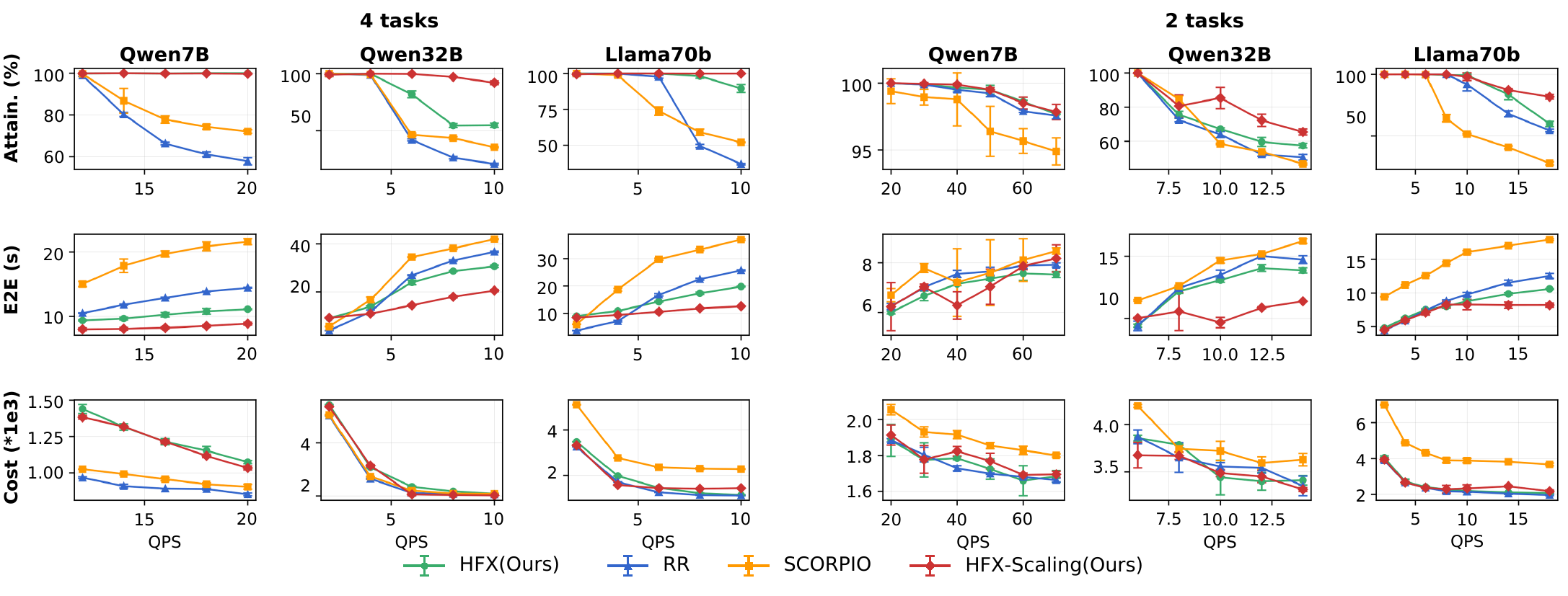}
    % \Description{Twenty-four line-graph figure showing multi-task performance for 4-task (leftmost three columns) and 2-task (rightmost three columns) workloads across Qwen7B, Qwen32B, and LLaMA70B models. The top row displays SLO attainment, with higher values indicating better performance. The middle row shows end-to-end latency, where lower values are better. The bottom row presents cost, also lower-is-better. Each graph compares four methods: \paper{}, RR, SCORPIO, and \paper{}-Scaling. Data is based on two instances per workload, with scaling experiments using up to four instances. The figure illustrates how the methods perform across different models and workloads, highlighting differences in SLO compliance, latency, and cost efficiency.}

    \caption{\textbf{Multi-task performance on 2-task and 4-task workloads.} Metrics include SLO attainment (top, higher is better), end-to-end latency (middle, lower is better), and cost (bottom, lower is better) for \paper{}, RR, SCORPIO, and \paper{}-Scaling. Results use two instances, with up to four for scaling.}

    % \caption{Multi-task performance on 2-task and 4-task workloads with Qwen7B, Qwen32B, and LLaMA70B. Metrics include SLO attainment (top, higher is better), end-to-end latency (middle, lower is better), and cost (bottom, lower is better) for \paper{}, RR, SCORPIO, and \paper{}-Scaling. Results use two instances, with up to four for scaling.}
    \label{fig:multitaskresult}
\end{figure*}

\textbf{Baselines.}
We compare \paper{} against four representative scheduling approaches. (1) \textsc{Aladdin}~\cite{nie2024aladdin}, a heuristic-based latency optimizer for single-task workloads; (2) \textsc{Simulated Annealing (SA)}~\cite{huang2025sloawareschedulinglargelanguage}, a stochastic scheduler designed for multi-SLO workloads; (3) \textsc{SCORPIO}~\cite{tang2025scorpioservingrightrequests}, which also targets multi-SLO workloads, extended with Round-Robin logic to support multi-instance scheduling; and (4) Llumnix~\cite{llumnix} with standard Round-Robin, referred as  \textsc{RR}.
Together, these baselines span both single-task and multi-SLO strategies, providing a comprehensive comparison across diverse competitive scenarios. All experiments are conducted under colocated deployment unless stated otherwise.

\textbf{Metrics.}
We evaluate three metrics (\S\ref{sec:problemformulation}): (i) SLO attainment, the fraction of requests meeting latency targets; (ii) E2E latency, from request arrival until request finishes generating tokens; and (iii) Resource cost, the total active time of all instances (one unit = one instance active for 50\,ms), capturing both instance count and duration for fair comparison across methods.

\subsection{Multi-task Workload Experiments}
To evaluate how \paper{} handles heterogeneous workloads, we test with two multi-SLO task sets under both deployment modes.
% Multi-task workloads with heterogeneous SLOs are common in production LLM deployments, where requests range from latency-sensitive chat to throughput-oriented batch summarization. To observe how the system handles such mixed workloads, we measure SLO attainment, end-to-end latency, and cost across 2-task and 4-task scenarios.

\subsubsection{Colocated Mode}

\begin{figure}[tbp]
    \centering
    \includegraphics[width=.78\linewidth]{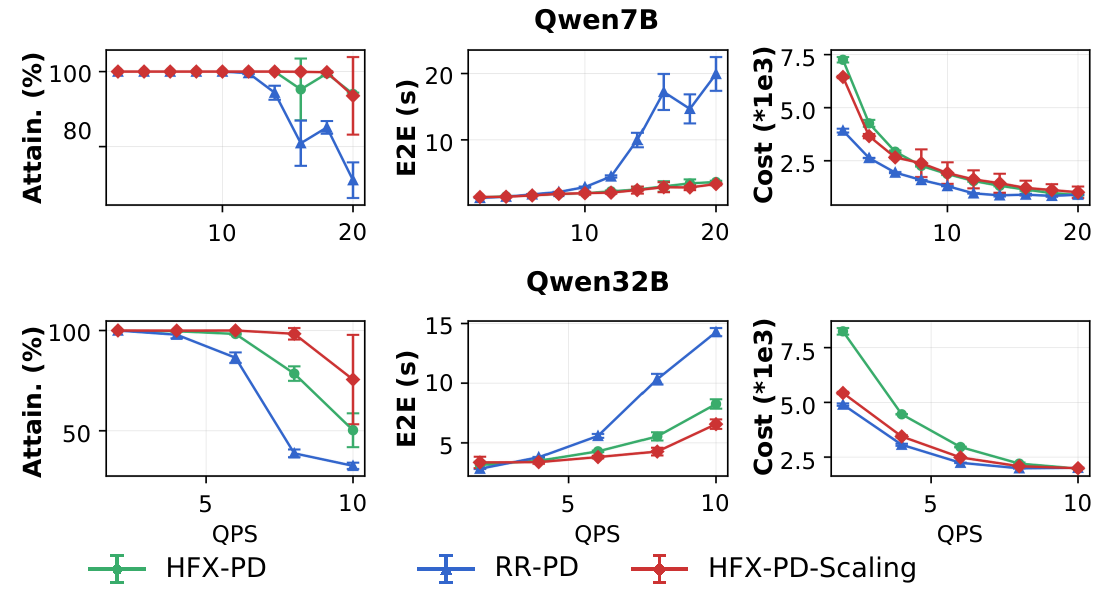}
    % \Description{Six line graphs arranged in a 2-by-3 grid. The top row shows results for the Qwen7B model, and the bottom row shows results for Qwen32B. The three columns correspond to SLO attainment (left), end-to-end latency (middle), and cost (right). Each graph plots performance over the 4-task workload under P/D execution. Different lines represent different methods being compared (\paper{}, RR, and \paper{}-PD-Scaling), showing how SLO attainment, latency, and cost vary for each model and method across the workload.}

    \caption{\textbf{Performance comparison under P/D disaggregated mode.} Results for 4-task workloads using Qwen7B and Qwen32B.}

    \label{fig:pd}
\end{figure}

Figure~\ref{fig:multitaskresult} compares the performance of \paper{} against \textsc{RR} and \textsc{SCORPIO}.
The top row reports \textbf{SLO attainment}.
On the 2-task benchmark, SLO-aware dispatching enables \paper{} to improve attainment by up to 1.23$\times$ over \textsc{RR} and 2.14$\times$ over \textsc{SCORPIO}.
With scaling enabled, \paper{}-Scaling further raises these gains to 1.5$\times$ and 2.93$\times$, respectively.
On the more heterogeneous 4-task benchmark, contention and queuing effects intensify, amplifying the benefits of SLO-aware scheduling.
\paper{} achieves up to 2.60$\times$ improvement over \textsc{RR} and 1.73$\times$ over \textsc{SCORPIO}, and \paper{}-Scaling further increases these improvements to 4.44$\times$ and 2.59$\times$.

The middle row reports \textbf{end-to-end latency}. While all methods show similar latency at low QPS, our advantages become clear as load increases, maintaining responsiveness under heavy traffic. On the 4-task benchmark, \paper{}-Scaling achieves up to 50.96\% and 65.82\% latency reduction over \textsc{RR} and \textsc{SCORPIO}, confirming that our SLO-aware policies effectively mitigate queuing and contention.

The bottom row compares \textbf{cost efficiency}. When compared to \textsc{RR}, our methods \paper{} and \paper{}-Scaling achieve similar cost, with occasional reductions of up to 4.99\%. In contrast, relative to \textsc{SCORPIO}, our methods deliver substantial cost savings, indicating that the improved user satisfaction of our methods does not come at the expense of higher costs. Notably, for Llama70B, we observe up to 49.81\% cost reduction over \textsc{SCORPIO}.

\subsubsection{P/D Disaggregated Mode}

% \sout{In many production deployments, prefill and decode stages run on separate pools to better utilize compute and memory resources. To evaluate this scenario, \paper{}-PD-Scaling,
% }
We denote our methods in the disaggregated deployment mode as \paper{}-PD (with the SLO-aware dispatcher) and \paper{}-PD-Scaling (with both the SLO-aware dispatcher and scaler). We test on 4-task benchmark with initial allocation of 2 prefill and 2 decode instances, and scaling up to 8 instances on demand.  Note that the experiment for Qwen32B (TP=2) runs across nodes, highlighting the system’s scalability and flexibility. Figure~\ref{fig:pd} compares \paper{} against RR-PD, showing that independent scheduling of prefill and decode improves SLO attainment and reduces latency while keeping cost similar to the baseline. The RR-PD baseline means that request dispatching in both the Prefill and Decode stages follows a round-robin policy. As shown in the Figure \ref{fig:pd}, both \paper{}-PD and \paper{}-PD-Scaling achieve higher attainment (up to 2.54$\times$) and lower request latency (up to 31.82\% reduction) compared to RR-PD. With auto-scaling enabled, the cost of \paper{}-PD-Scaling is only slightly higher than RR-PD for Qwen7B and roughly equivalent to RR-PD for Qwen32B. These experiments demonstrate practical benefits of our methods under disaggregated mode and heterogeneous workloads.

\begin{figure}[tbp]
    \centering
    \includegraphics[width=.9\linewidth]{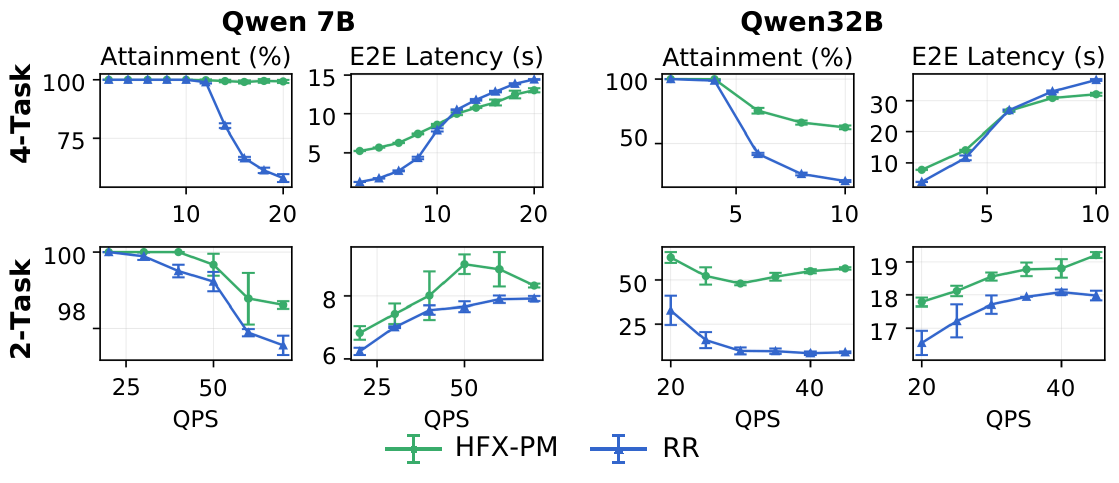}
    % \Description{Eight line graphs arranged in a 2-by-4 grid. The top row shows results for 4-task workloads, and the bottom row shows 2-task workloads. The left two columns correspond to the Qwen7B model, and the right two columns correspond to the Qwen32B model. Within each model, the left graph shows SLO attainment and the right graph shows end-to-end latency. Each graph compares performance of \paper{} against RR across dynamic SLO ranges and varying QPS, showing that \paper{} consistently achieves higher SLO attainment, especially at higher QPS.}
    % \caption{\textbf{Performance of \paper{}-PM under dynamic SLO ranges.} Results for 2-task and 4-task workloads using Qwen7B and Qwen32B. Metrics shown are SLO attainment and end-to-end latency comparing \paper{}-PM against RR. \paper{}-PM consistently achieves higher SLO attainment, especially at higher QPS.}
    \caption{Performance under dynamic SLO ranges for Qwen7B and Qwen32B across 2-task and 4-task workloads. Metrics shown are SLO attainment and end-to-end latency comparing \paper{} against RR. \paper{} consistently achieves higher SLO attainment, especially at higher QPS.}
    \label{fig:priortiyresult}
\end{figure}

\subsection{Priority Mapping Experiments}

In production, requests often carry different latency priorities. To evaluate whether \paper{} can handle dynamic SLO assignment based on priority, we use median SLO targets with a ±25\% range to separate priorities. Figure~\ref{fig:priortiyresult} shows that \paper{} consistently outperforms \textsc{RR} in SLO attainment across all workloads, with the advantage increasing at higher QPS due to its multi-SLO-aware scheduling and predictive mechanisms, achieving up to 7.02$\times$ improvement in attainment. Latency varies with workload and load: for 2-task sets, \paper{} incurs slightly higher latency due to scheduling overhead and conservative resource allocation with at most 1.35s, whereas for 4-task sets at high QPS, it reduces latency through efficient dispatching, prioritization, and lower queuing delays. The cost metric, which reflects scaling overhead, is not shown here. These results reflect the system’s ability to honor customer-assigned priorities under varying load, a common operational requirement.

\subsection{Fast Scaling Experiments}
Workloads fluctuate in practice, requiring the system to rapidly scale without violating SLOs. We compare three scaling strategies: (1) loading weights from disk when scaling, (2) offloading weights to CPU when scaling out and reloading from CPU when scaling in, and (3) loading weights from an already running instance, which we term \emph{Fast Scaling}. As shown in Table~\ref{tab:scaletime}, Fast Scaling significantly reduces scaling time, achieving up to a 9.88$\times$ speedup over CPU offloading and 19.39$\times$ speedup over disk-based loading. This shows that reusing weights resident in peer device memory can dramatically accelerate dynamic scaling without incurring the overhead of disk or CPU transfers.

\begin{table}[tbp]
\setlength{\tabcolsep}{3pt}
\centering
\begin{tabular}{llll}
\toprule
               & \multicolumn{1}{c}{Qwen 7B} & \multicolumn{1}{c}{Qwen 32B} & \multicolumn{1}{c}{Llama 70B} \\
\bottomrule
Fast Scaling & \textbf{0.89 $\pm$ 0.01} & \textbf{2.05 $\pm$ 0.02} & \textbf{1.16 $\pm$ 0.05}\\
CPU Offloading & 2.73 $\pm$ 0.17 & 19.41 $\pm$ 3.14 & 11.50 $\pm$ 1.86\\
Disk loading   & 4.14 $\pm$ 0.26 & 28.84 $\pm$ 2.23 & 22.58 $\pm$ 0.82 \\    \bottomrule      
\end{tabular}
\caption{\textbf{Scaling time for different models.} Results for Qwen7B, Qwen32B, and LLaMA70B using different methods, measured in seconds.}

% \caption{Scaling time (s) for Qwen7B, Qwen32B, and Llama70B using different methods}
\label{tab:scaletime}
\end{table}

%-------------------------------------------------------------------------------

%-------------------------------------------------------------------------------
\subsection{Deployment Experiences}
\label{sec:ablation}

We evaluate how \paper{} behaves under dynamic workloads and varying system parameters, highlighting insights relevant for production deployment. These experiments do not aim to quantify algorithmic contributions, but to demonstrate practical robustness and responsiveness in realistic settings.

\subsubsection{Dynamic SLO Assignment and Priority Mapping}
% \zhenan{move to ablation}

Production LLM deployments experience fluctuating request rates and heterogeneous priority levels. To evaluate our system's ability to dynamically adapt to runtime conditions by mapping priorities to absolute SLOs (Section~\ref{sec:pmap}), we conduct experiments under a dynamic workload.

\begin{figure}[t]
    \centering
    \includegraphics[width=\linewidth]{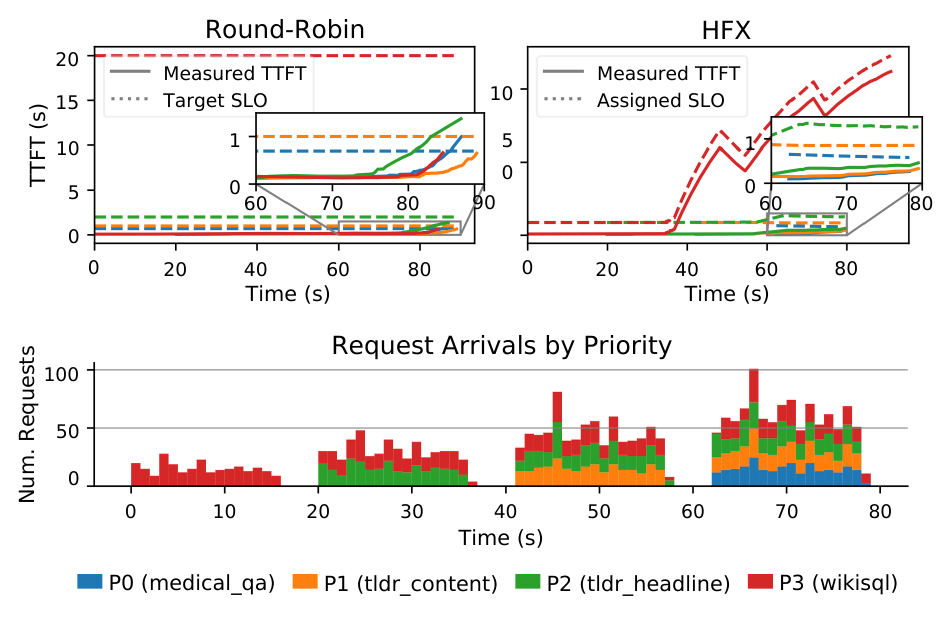}
    % \Description{Three-panel figure illustrating \paper{}’s responsiveness to dynamic workloads. The top row has two line graphs: the left shows TTFT using Round Robin (RR), and the right shows TTFT using \paper{}. The bottom row shows a stacked histogram of request arrivals by priority, starting with lowest-priority tasks (P3) at 15 QPS from a single client, with higher-priority tasks (P2, P1, P0) gradually joining until all four priorities are active at a total rate of 60 QPS after 60 seconds. The line graphs illustrate that RR schedules requests as quickly as possible, causing higher-priority SLO violations under contention, whereas \paper{} dynamically adjusts SLO assignments, reducing latency for low-priority tasks when underloaded and preserving capacity for high-priority tasks under contention, maintaining all task SLOs.}
    \caption{\textbf{Responsiveness of \paper{} to dynamic workloads.} Top: The measured TTFT  (solid lines) and the TTFT SLO (dotted lines) for different request types, averaged over per-second intervals for the RR (left) and \paper{} (right) schedulers. Bottom: The request arrival rate, where P3 starts at 15 QPS and priorities P2–P0 gradually ramp up to a total of 60 QPS at 60 seconds. 
    % RR schedules greedily, causing high-priority SLO violations under contention, while \paper{} dynamically adjusts SLOs—lowering latency for low-priority tasks when underloaded and preserving capacity for high-priority tasks under contention, maintaining all SLOs.
    % The top row shows TTFT with RR (left) and \paper{} (right). The bottom row plots request arrivals by priority: P3 starts at 15 QPS, and P2–P0 gradually join until reaching 60 QPS after 60s. RR schedules greedily, causing high-priority SLO violations under contention, while \paper{} dynamically adjusts SLOs—lowering latency for low-priority tasks when underloaded and preserving capacity for high-priority tasks under contention, maintaining all SLOs.
    }
    \label{fig:dynamic_slo}
\end{figure}

% \textbf{Dynamic Task Rate}
We begin with a single client generating 250 requests lowest-priority tasks (P3 tasks) at 15 QPS. After 20 seconds, a new client joins, issuing P2 tasks at the same rate. This process continues until all four priority levels are active, at which point the total task rate reaches 60 QPS after 60 seconds.
The bottom row of Figure~\ref{fig:dynamic_slo} shows the resulting stacked request histogram.

In this experiment, absolute task SLOs are derived from Table~\ref{tab:task-stats}, and \paper{}’s priority mapping is configured to satisfy these targets. 
% We show TTFT results to study the behaviour of the system and omit the TPOT results.

As shown in the upper row of Figure~\ref{fig:dynamic_slo}, \textsc{RR} schedules requests as quickly as possible, creating contention when multiple task types coexist (60–90s window), causing the TTFT of higher-priority tasks to violate their SLOs (blue dotted line).

In contrast, \paper{} dynamically adjusts its scheduling. When the system is underloaded (first 20s), it assigns tighter SLOs to low-priority (P3) requests to minimize their latency. As higher-priority requests join and contention increases, \paper{} adapts by relaxing the SLOs for lower-priority tasks up to their configured maximum, thereby preserving capacity to meet the strict SLOs of higher-priority tasks.

The results demonstrate that \paper{} can safely handle sudden changes in demand while preserving service guarantees, a critical property for operational deployments. By dynamically adjusting assigned SLOs, it optimizes latency when capacity is available and ensures strict prioritization under contention, ensuring all tasks meet their configured performance targets.

\subsubsection{Single-task Stability}
% Even though \paper{} is designed for multi-task, multi-SLO workloads, single-task deployments remain common in production. 
% To evaluate \paper{} in a single-task setting, 
\paper{} is designed for multi-task, multi-SLO workloads, we want to verify that it is still working on single-task deployments that remain common in production. 
We evaluate on the WikiSQL dataset using Qwen7B and Qwen32B, and compare \paper{} against two state-of-the-art single-task, single-SLO baselines, \textsc{Aladdin} and \textsc{SA}, under varying QPS levels. For this experiment, the TTFT and TPOT SLOs are set to 0.7 s and 0.5 s, respectively.
% to stress the systems under tight latency requirements.

\begin{figure}
    \centering
    \includegraphics[width=.66\linewidth]{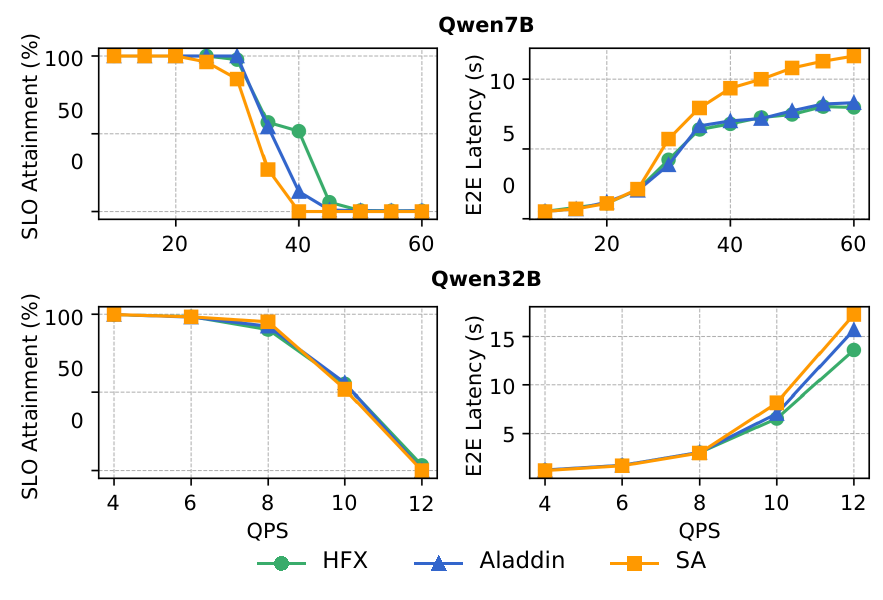}
    % \Description{Four line graphs arranged in a 2-by-2 grid. The top row shows results for the Qwen7B model, and the bottom row shows results for Qwen32B. The left column displays SLO attainment, and the right column displays end-to-end latency. Each graph shows single-task performance on the WikiSQL dataset under TTFT and TPOT SLOs set to 0.7 seconds and 0.5 seconds, respectively. Different lines in each graph represent the methods (\paper{}, Aladdin and SA) being compared, illustrating their performance for each model and metric.}
    % \caption{\textbf{Single-task performance comparison.} Results for Qwen7B and Qwen32B on the WikiSQL dataset in compare with Aladdin~\cite{nie2024aladdin} and SA~\cite{huang2025sloawareschedulinglargelanguage}. The TTFT and TPOT SLOs are set to 0.7\,s and 0.5\,s, respectively.}

    \caption{Single-task performance comparison of Qwen7B and Qwen32B on the WikiSQL dataset. The TTFT and TPOT SLOs are set to 0.7\,s and 0.5\,s, respectively.}
    \label{fig:singletaskresult}
\end{figure}

As shown in Figure~\ref{fig:singletaskresult}, all systems maintain near-perfect SLO attainment under light load. As QPS increases, however, \textsc{SA}’s attainment drops sharply once the system approaches saturation, while \paper{} sustains higher attainment, particularly at the knee point for Qwen7B (40~QPS). 

% At low to moderate QPS, all systems exhibit similar latencies. 
Figure~\ref{fig:singletaskresult} also reports end-to-end latency. Under higher loads, however, \paper{} delivers more stable performance, avoiding the sharp latency spikes observed with \textsc{SA} and \textsc{Aladdin} (e.g., at 12~QPS on Qwen32B). These results confirm that \paper{}’s scheduling and instance-aware queuing mechanisms improve responsiveness and reduce congestion even in single-task settings.

Overall, these results show that although \paper{} targets heterogeneous multi-SLO workloads, it remains competitive, and sometimes even better than \textsc{Aladdin} and \textsc{SA}.
% Overall, these results show that although \paper{} is designed to handle heterogeneous multi-SLO workloads, it remains competitive with, and in some cases outperforms, state-of-the-art methods \textsc{Aladdin} and \textsc{SA} in single-task settings. 
% That is, the generality of \paper{} does not come at the expense of performance on simpler workloads.

\subsubsection{Monitor and Scaling Intervals}

To understand the sensitivity of \paper{}’s performance to key system parameters, we conduct an ablation study varying both the monitoring interval and the scaling interval as shown in Figure~\ref{fig:ablation}.

\textbf{Effect of Monitor Interval.} We vary the monitor interval across 50\,ms, 1\,s, and 5\,s. As shown in the top row of Figure~\ref{fig:ablation}, performance is largely robust to the choice of interval. However, slight trends are observable: (1) SLO attainment is slightly reduced with a 5\,s monitor interval compared to 50\,ms and 1\,s, reflecting delayed feedback that slows adaptation to system status changes; (2) end-to-end latency is marginally higher with the 50\,ms interval due to the slight overhead of frequent monitoring, especially under heavy load, though the impact remains minimal. Overall, \paper{} is robust across a wide range of monitor intervals, and choosing a moderate interval (e.g., 1\,s) provides a good balance between responsiveness and overhead.

\textbf{Effect of Scaling Interval.} We vary the scaling interval across 0.5\,s, 1\,s, and 2\,s to evaluate how the frequency of scaling decisions affects performance (Figure~\ref{fig:ablation}). SLO attainment shows minimal differences across intervals for most QPS values. At high QPS, shorter intervals exhibit slightly lower attainment, as the overhead of frequent scaling outweighs the benefits, particularly under heavy load when time is critical.  

These results suggest that \paper{}’s performance is largely insensitive to the scaling interval within the tested range. While very short or intermediate intervals may introduce minor variability in attainment or latency, overall system robustness is preserved. These insights guide deployment parameter choices without requiring deep algorithmic tuning.

\begin{figure}
    \centering
    \includegraphics[width=.66\linewidth]{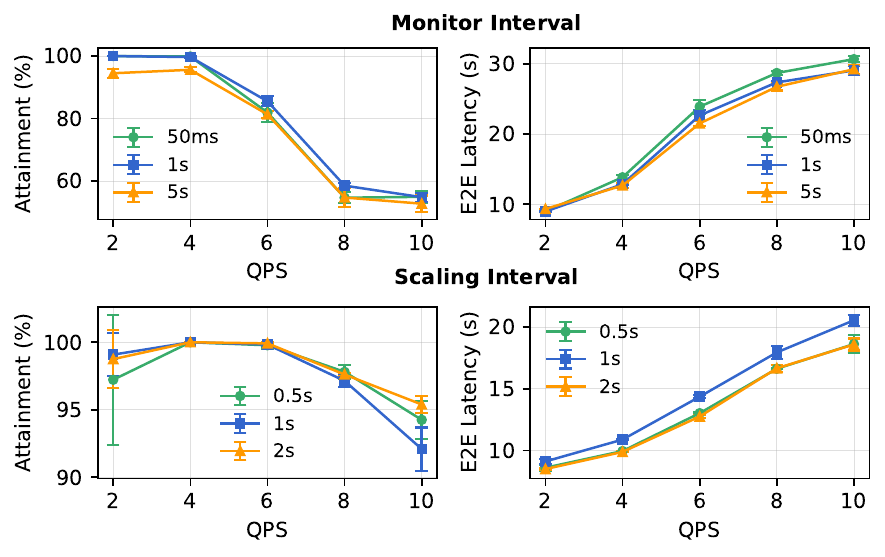}
    % \Description{Four line graphs arranged in a 2-by-2 grid. The top row shows results for different monitor intervals, and the bottom row shows results for different scaling intervals. The left column displays SLO attainment, and the right column displays end-to-end latency. The first row uses two instances, and the second row uses up to four instances. Each graph illustrates the impact of varying monitor or scaling intervals on Qwen32B’s performance under the 4-task benchmark, comparing how these settings affect SLO attainment and latency.}
    % \caption{\textbf{Impact of monitor and scaling intervals.} Results for Qwen32B on the 4-task benchmark are shown with two instances (top) and up to four instances (bottom). Overall, \paper{}’s performance is largely insensitive to changes in the scaling and monitor intervals.}
    \caption{Ablation of monitor and scaling intervals on Qwen32B with the 4-task benchmark, using two instances in the first row and up to four instances in the second row.}
    \label{fig:ablation}
\end{figure}

%-------------------------------------------------------------------------------

%-------------------------------------------------------------------------------
\section{Related Work}
\label{sec:related_works}

\subsection{Single-Node Inference and Prefill/Decode Optimization}
Single-node LLM serving systems have focused on improving efficiency within one server. vLLM~\cite{kwon2023efficient} and FlexGen~\cite{sheng2023flexgen} optimize KV reuse and CPU–GPU offloading to reduce memory footprint, but assume uniform workloads without SLO differentiation. Apt-Serve~\cite{10.1145/3725394} and Sarathi-Serve~\cite{agrawal2024taming} improve throughput and tail-latency control via adaptive batching and chunked-prefill scheduling.

These systems do not support multi-SLO-aware scheduling or dynamic scaling. In production, heterogeneous workloads with conflicting latency and throughput targets can lead to SLO violations or inefficient utilization when relying on single-node optimizations alone. This motivates the need for a multi-SLO-aware dispatcher that works across instances and stages, not just within a single node.

\subsection{Cluster-Level LLM Serving}
Distributed LLM serving frameworks exploit multiple nodes to handle large models and diverse workloads. Disaggregated deployment such as DistServe~\cite{zhong2024distserve} and Splitwise~\cite{patel2024splitwiseefficientgenerativellm} separate prefill and decode onto specialized cluster pools to balance compute and memory usage. KV-cache optimization systems like MIRAGE~\cite{li2025miragekvcacheoptimization} and BlitzScale~\cite{zhang2025blitzscale} improve memory utilization and tail latency through remapping and layer-level autoscaling.

While these approaches improve resource utilization and latency, they often assume static placement and simple scheduling policies. In real-world cloud deployments, workload intensity can fluctuate rapidly, and prefill/decode stages may require independent scaling. These limitations motivate dynamic stage-aware dispatching and adaptive scaling mechanisms to maintain SLOs under heterogeneous and bursty workloads.

\subsection{Multi-SLO Scheduling and Scaling}
Recent work addresses scheduling and scaling across multiple SLOs. SCORPIO~\cite{tang2025scorpioservingrightrequests}, Tempo~\cite{zhang2025tempoapplicationawarellmserving}, and SA~\cite{huang2025sloawareschedulinglargelanguage} prioritize requests based on SLO type and input length. Arrow~\cite{wu2025arrowadaptiveschedulingmechanisms}, Aladdin~\cite{nie2024aladdin}, and PD-Multiplexing~\cite{cui2025optimizingsloorientedllmserving} extend this to prompt/decode architectures, allocating workloads dynamically to reduce latency or improve throughput. PolyServe~\cite{zhu2025polyserveefficientmultisloserving} and SLOs-Serve~\cite{chen2025slosserveoptimizedservingmultislo} incorporate multi-SLO constraints into placement. HAS-GPU~\cite{gu2025hasgpuefficienthybridautoscaling} and USHER~\cite{298766} introduce adaptive autoscaling across replicas.
% \textcolor{green}{maybe - > ignore the slow scaling time}. 
These systems often treat scheduling, placement, and scaling separately, or assume slow spin-up times. They do not proactively reallocate capacity during bursty multi-SLO workloads, or limitedly support P/D disaggregation. In operational environments, this can lead to SLO violations and underutilized resources. These gaps motivate a unified system that jointly manages multi-SLO-aware scheduling, P/D disaggregation, and fast adaptive scaling.

\section{Conclusion}
\label{sec:conclusion}
We presented \paper{}, a unified LLM serving system that addresses the challenges of multi-SLO scheduling and elastic scaling. By combining a multi-SLO-aware dispatcher, a two-stage P/D scheduling mechanism, adaptive SLO mapping, and a fast, unified scaling controller, \paper{} efficiently manages heterogeneous workloads across both colocated and disaggregated deployments. It also supports distributed settings, including cross-node scenarios. Our extensive evaluation demonstrates that \paper{} significantly improves SLO attainment, reduces latency, lowers operational cost, and accelerates scaling compared to existing baselines. These results show that by integrating SLO-aware dispatching with dynamic scaling, \paper{} provides an effective, high-performance, and cost-efficient solution for LLM serving across diverse, multi-task workloads, generalizing across workload sizes, model scales, and deployment architectures to support production-grade, multi-tenant inference systems.

Future work includes exploring heterogeneous models and hardware, enabling \paper{} to efficiently serve multiple model types of different sizes and resource requirements on both low-end and high-end hardware simultaneously.
%-------------------------------------------------------------------------------

\newpage
% \bibliographystyle{ACM-Reference-Format}
% \bibliography{asplos/references}
\bibliographystyle{plain}
\bibliography{references}

% \newpage
% \appendix

% \end{CJK*}
\end{document}

%% file: Meta/package.tex
\usepackage{subcaption}
\usepackage[sort]{natbib}
\usepackage{dblfloatfix}
\usepackage{ulem}
\usepackage{float}
\usepackage{caption}
\usepackage{dramatist}
\usepackage{xspace}
\usepackage{pifont} % http://ctan.org/pkg/pifont
\usepackage{multirow}
\usepackage{tcolorbox}
\usepackage{xltabular}
\usepackage{longtable}
\usepackage{mdframed}
\usepackage{hyperref}
\usepackage{tocloft}
\usepackage{amsfonts}
\usepackage{amsmath}
\usepackage{amssymb}
\usepackage{lineno}
\usepackage{multirow}
\usepackage{adjustbox}

\usepackage[bottom]{footmisc}
\usepackage{CJKutf8}
\usepackage{setspace}

\usepackage{graphicx}

\usepackage{dsfont}
\usepackage{array} % 用于tabularx环境
\usepackage{tabularx} % 用于tabularx环境
\usepackage{xcolor} % 用于画有颜色的框

\usepackage{lipsum}  % 用于生成示例文本
\usepackage{multicol} % 引入 multicol 宏包

\makeatletter
\def\@BTrule[#1]{%
  \ifx\longtable\undefined
    \let\@BTswitch\@BTnormal
  \else\ifx\hline\LT@hline
    \nobreak
    \let\@BTswitch\@BLTrule
  \else
     \let\@BTswitch\@BTnormal
  \fi\fi
  \global\@thisrulewidth=#1\relax
  \ifnum\@thisruleclass=\tw@\vskip\@aboverulesep\else
  \ifnum\@lastruleclass=\z@\vskip\@aboverulesep\else
  \ifnum\@lastruleclass=\@ne\vskip\doublerulesep\fi\fi\fi
  \@BTswitch}
\makeatother

\addto\extrasenglish{
}

 {\begin{list}{}%
         {\setlength{\leftmargin}{#1}}%
         \item[]%
 }
 {\end{list}}

\reportnumber{001} % Leave blank if n/a
\renewcommand{\today}{}

%% file: Meta/macro.tex
\usepackage{tikz}